\documentclass[]{piparticle-final}
\usepackage{graphicx}
\usepackage{amsmath}
\usepackage{textcomp}

\begin{document}

\volume{2}               
\articlenumber{020001}   
\journalyear{2010}       
\editor{S. A. Cannas}   
\reviewers{P. Netz, Inst. de Qu\'{\i}mica, Univ. Federal do Rio Grande do Sul, Brazil}  
\received{13 July 2009, Revised: 29 December 2009}     
\accepted{7 February 2010}   
\runningauthor{M. G. Campo}  
\doi{020001}         

\title{Structural and dynamic properties of SPC/E water}

\author{M. G. Campo,\cite{inst1}\thanks{E-mail: mario@exactas.unlpam.edu.ar}
        }

\pipabstract{

  I have investigated the structural and dynamic properties of water by performing a series of molecular dynamic simulations in the range of temperatures from 213 K to 360 K, using the Simple Point Charge-Extended (SPC/E) model. I performed isobaric-isothermal simulations (1 bar) of 1185 water molecules  using the GROMACS package. I quantified the structural properties using the oxygen-oxygen radial distribution functions, order parameters, and the hydrogen bond distribution functions, whereas, to analyze the dynamic properties I studied the behavior of the history-dependent bond correlation functions and the non-Gaussian parameter $\alpha_{2}(t)$ of the mean square displacement of water molecules. When the temperature decreases, the translational ($\tau$) and orientational (\textit{Q}) order parameters are linearly correlated, and both increase indicating an increasing structural order in the systems. The probability of occurrence of four hydrogen bonds and \textit{Q} both have a reciprocal dependence with $T$, though the analysis of the hydrogen bond distributions permits to describe the changes in the dynamics and structure of water more reliably. Thus, an increase on the \textit{caging} effect and the occurrence of long-time hydrogen bonds occur below $\sim$ 293 K, in the range of temperatures in which predominates a four hydrogen bond structure in the system.

}

\maketitle
\blfootnote{
\begin{theaffiliation}{99}
   \institution{inst1} Universidad Nacional de La Pampa, Facultad de Ciencias Exactas y Naturales, Uruguay 151, 6300 Santa Rosa, La Pampa, Argentina.
\end{theaffiliation}
}

\section{Introduction}

  Water is the subject of numerous studies due to its biological significance and its universal presence [1--3]
. The thermodynamic behavior of water presents important differences compared with those of the other substances, and many of the characteristics of such behavior are often attributed to the existence of hydrogen bonds between water molecules. Scientists have found that the water structure produced by the hydrogen bonds is peculiar as compared to that of other liquids. Then, the advances in the knowledge of hydrogen bond behavior are crucial to understanding water properties.

The method of molecular dynamics (MD) allows to analyze the structure and dynamics of water at the microscopic level and hence to complement experimental techniques in which these properties can be interpreted only in a qualitative way (infra-red absorption and Raman scattering \cite{ref04}, depolarized light scattering \cite{ref05,ref06}, neutron scattering \cite{ref07}, femtosecond spectroscopy [8--11] 
and other techniques [12--14]
.

Among the usual methods to study the short range order in MD simulations of water are the calculus of radial distribution functions, hydrogen bond
distributions and order parameters. The orientational order
parameter \textit{Q} measures the tendency of the system to adopt a
tetrahedral configuration considering the water oxygen atom as
vertices of a tetrahedron, whereas the translational order parameter
$\tau $ quantifies the deviation of the pair correlation function
from the uniform value of unity seen in an ideal gas \cite
{ref15,ref16}. The order parameters are used to construct an order
map, in which different states of a system are mapped onto a plane
$\tau $-\textit{Q}. The order parameters are, in general, independent,
but they are linearly correlated in the region in which the water
behaves anomalously \cite{ref16a}.

The dynamics of water can by characterized by the bond lifetime, $\tau _{HB}$, associated to the process of rupturing and forming of hydrogen bonds between water molecules which occurs at very short time scale [9, 18, 19, 21--23]
. $\tau _{HB}$ is obtained in MD using the history-dependent bond correlation function $P(t)$, which represents the probability that an hydrogen bond formed at time $t=0$ remained continuously unbroken and breaks at time $t$  \cite{ref23,ref24}.

Also, the dynamics of water can be studied by analyzing
the mean-square displacement time series $M(t)$. In addition to the diffusion coefficient calculation at long times in which $M(t)\propto t$, in the supercooled region of temperatures and at intermediate times $M(t)\propto t^{\alpha}$ ($0<\alpha <1$). This behavior of $M(t)$ is associated to the subdiffusive movement of the water molecules, caused by the \textit{caging} effect in which a water molecule is temporarily trapped by its neighbors and then moves in short bursts due to nearby cooperative motion. A time $t^{\ast}$ characterizes this \textit{caging} effect (see Sec. II for more details) \cite{ref24a,ref24b}.

In a previous work, we found a $q$-exponential behavior in $P(t)$, in which $q$ increases with $T^{-1}$ approximately below 300 K. $q(T)$ is also correlated with the probability of occurrence of four hydrogen bonds, and the subdiffusive motion of the water molecules \cite{ref25}.

The relationship between dynamics and structural properties of water has not been clearly established to date. In this paper, I explore whether the effect that temperature has on the water dynamics reflects a more general connection between the structure and the dynamics of this substance.

\section{Theory and method}

  I have performed molecular dynamic simulations of SPC/E water model using the GROMACS package \cite{ref26,ref27}, simulating fourteen similar systems of 1185 molecules at 1 bar of pressure in a range of temperatures from 213 K to 360 K. I initialized the system at 360 K using an aleatory configuration of water molecules, assigning velocities to the molecules according to a Boltzmann's distribution at this temperature. For stabilization, I applied Berendsen's thermal and hydrostatic baths at the same temperature and 1 bar of pressure \cite{ref28}. Then, I ran an additional MD obtaining an isobaric-isothermal ensemble. I obtained the other systems in a similar procedure, but using as initial configuration that of the system of the preceding higher temperature and cooling it at the slow rate of 30 K ns$^{-1}$ \cite{ref16a}. Stabilization and sampling periods for the systems at different temperatures are indicated in Table \ref{tab01}. Simulation and sampling time steps were 2 fs and 10 fs, respectively. The sampling time step was shorter than the typical time during which a hydrogen bond can be destroyed by libration movements.

\begin{table}[tbp]
\begin{center}
\caption{Details of the simulation procedure. Duration of the stabilization period ($t_{est}$) and the MD sampling ($t_{MD}$) in the different ranges of temperatures}
\label{tab01}%
\begin{tabular}{c|cc}
\hline\hline Temp. range (K) & $t_{est}$ (ns) & $t_{MD}$ (ns) \\ 
\hline 213 - 243 & 20.0 & 10.0 \\ 
253 - 273 & 16.0 & 10.0 \\ 
283 - 360 & 16.0 &  8.0 \\ \hline\hline
\end{tabular}%
\end{center}
\end{table}

\begin{figure}[h]
\begin{center}
\includegraphics*[width=7.4cm]{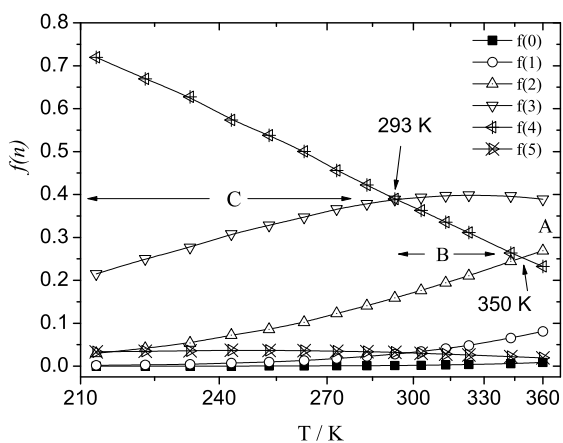}
\end{center}
\caption{Hydrogen bonds distribution functions $f(n)$ ($n=0,...,5$) versus $T$. The zones A, B and C correspond to ranges of temperatures in which occur different relationships between $f(4), f(3)$ and $f(2)$. Note the reciprocal scale for the temperatures. See the text for details.}
\label{figure01}
\end{figure}
I calculated the hydrogen bond distribution functions $f(n)$ ($n=0,1,...,5$), which is the probability of occurrence of $n$ hydrogen bonds by molecule, considering a geometric definition of hydrogen bond \cite{refhb01}. As parameters for this calculation, I used a maximum distance between oxygen atoms of 3.5 $\mathring{A}$ and a minimum angle between the atoms O$_{donor}$--H--O$_{acceptor}$ of 145$^{\circ}$.

The radial distribution function (RDF) is a standard tool used in experiments, theories, and simulations to characterize the structure of condensed matter. Using RDFs, I obtained the average number, $N$, of water molecules in the first hydration layer (the hydration number)

\begin{equation}
N= 4 \pi \rho \int_{0} ^{r_{min}} g(r) r^{2} dr
\label{ec01}
\end{equation}%

where $\rho$ is the number density. 
\begin{figure}[h]
\begin{center}
\includegraphics*[width=7.4cm]{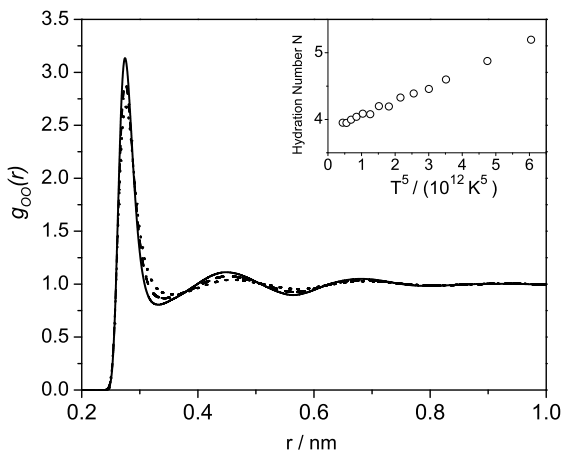}
\end{center}
\caption{Oxygen-oxygen radial distribution functions for the systems
at 213 K (continuous line), 293 K (dashed line), and 360 K (dotted line).
Inset: The hydration number $N$ vs. $T^{5}$.}
\label{figure02}
\end{figure}

The translational order parameter, $\tau$, is defined in Ref. \cite{ref16} as

\begin{equation}
\tau\equiv\int_{0} ^{S_{c}} |g(s)-1|ds
\label{ec02}
\end{equation}%
where the dimensionless variable $s\equiv rn^{1/3}$ is the radial distance r scaled by the mean intermolecular distance $n^{1/3}$, and $S_{c}$  corresponds to half of the simulation box size.

The orientational order parameter $Q$ is defined as \cite{ref15}
\begin{equation}
Q= \left \langle 1- \frac{3}{8} \sum_{i=1}^N \sum_{j=1}^4 \sum_{l=j+1}^4 \left [ cos\theta_{jik} + \frac{1}{3} \right ]^2 \right \rangle
\label{ec03}
\end{equation}%
where $\theta_{jik}$ is the angle formed by the atoms O$_{j}$--O$_{i}$--O$_{k}$. Here, O$_{i}$ is the reference oxygen atom, and O$_{j}$ and O$_{k}$ are two of its four nearest neighbors. $Q$=1 in an ideal configuration in which the oxygen atoms would be located in the vertices of a tetrahedron.

I obtained the bond correlation function $P(t)$ from the simulations by building a histogram of the hydrogen bonds lifetimes for each configuration. Then, I fitted this function with a Tsallis distribution of the form

\begin{equation}
\exp _{q}(t)=\left[ 1+\left( 1-q\right) t\right] ^{1/\left( 1-q\right) }
\label{ec04}
\end{equation}%
being $t$ the hydrogen bond lifetime and $q$ the nonextensivity parameter \cite{ref25,ref29}. If $q = 1$, Eq. (\ref{ec04}) reduces to an exponential, whereas if $q > 1$, $P(t)$ decays more slowly than an exponential. This last behavior occurs when long lasting hydrogen bonds increase their frequency of occurrence. 

The subdiffusive movement of water occurs when the displacement of the molecules obeys a non-Gaussian statistics. This behavior is characterized by $t^{*}$, the time in which the non-Gaussian parameter $\alpha_{2}(t)$ reaches a maximum [see Eq.
(\ref{ec05})]. Then, $t^{*}$ is the parameter associated to the average time during which a water molecule is trapped by its environment (\textit{caging} effect), and this prevents it from reaching the diffusive state \cite{ref24a,ref24b}.

\begin{equation}
\alpha_{2}(t)=\frac{3 \langle r^{4}(t) \rangle}{5 \langle r^{2}(t) \rangle} -1
\label{ec05}
\end{equation}%

\section{Results and discussion}

  Three zones or ranges of temperatures can be distinguished in the
graph of the hydrogen bond distributions $f(n)$ vs. $T$ (see Fig.
\ref{figure01}). Zone A ($T$ $>$ 350 K) in which $
f(3)>f(2)>f(4)$, zone B (293 K $> T >$ 350 K) in which
$f(3)>f(4)>f(2)$, and zone C ($T$ $<$ 293 K) in which
$f(4)>f(3)>f(2)$. These results indicate a predominant structure of
three and two hydrogen bonds (3HB-2HB) in zone
A, 3HB-4HB in zone B, and 4HB-3HB in zone C,
respectively. $f(4)\propto T^{-1}$ in all ranges of temperatures, showing that the tetrahedral structure of water decreases with the increase of this variable.$f(3)$ increases with $T$ up to 293 K, and then remains approximately constant ($\sim 0.4$) up to 360 K. $f(2)$ also increases with ${T}$ in all range of temperatures, but only overcomes  $f(4)$ at $T >$ 350 K.

\begin{figure}[h]
\begin{center}
\includegraphics*[width=7.4cm]{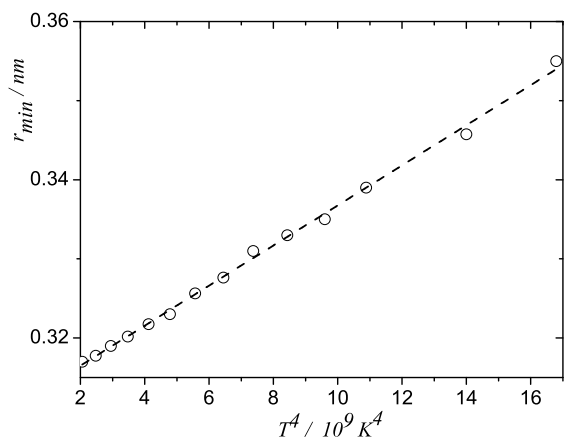}
\end{center}
\caption{ Position of the first minimum of the oxygen-oxygen radial
distribution function vs $T^{4}$, associated to the size of the first hydration layer.}
\label{rmin}
\end{figure}

Fig. \ref{figure02} shows the oxygen-oxygen RDFs corresponding to the systems at 213 K, 293 K and 360 K. When the temperature decreases, the minimum and maximum tend to be more defined. This being associated with an increasing order in the system. The position of the first minimum moves closer to the origin decreasing the size of the first hydration layer ($\propto T^{4}$ see Fig. \ref{rmin}). Both facts can be associated to the decrease of the hydration number from $N \sim$ 5 to $N \sim$ 4 (see inset, Fig. \ref{figure02}).

\begin{figure}[h]
\begin{center}
\includegraphics*[width=7.4cm]{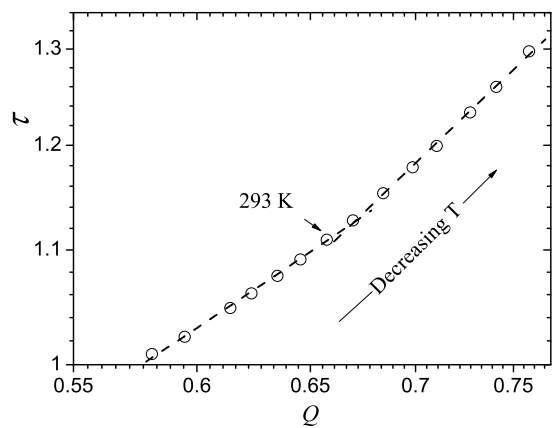}
\end{center}
\caption{Order map with the values of the order parameters
corresponding to the simulated systems. Note the change in the
slope of the line at $T\sim$ 273 K} \label{figure03}
\end{figure}

The simultaneous behavior of \textit{Q} and $\tau$ is shown in the order map of Fig. \ref{figure03}, in which the location of the values corresponding to 293 K are indicated by an arrow. The order parameters present similar behaviors with the temperature. Upon cooling, these parameters are linearly correlated and move in the order map along a line of increasing values, up to reaching maximum values at 213 K. The slope of the line increases a little for $T >$ 293 K, indicating that $\tau$ has a response to the increase of $T$ slightly higher than \textit{Q} in this range of temperatures. The positive values of the slopes indicate an increasing order of the system when the temperature decrease.

\begin{figure}[h]
\begin{center}
\includegraphics*[width=7.4cm]{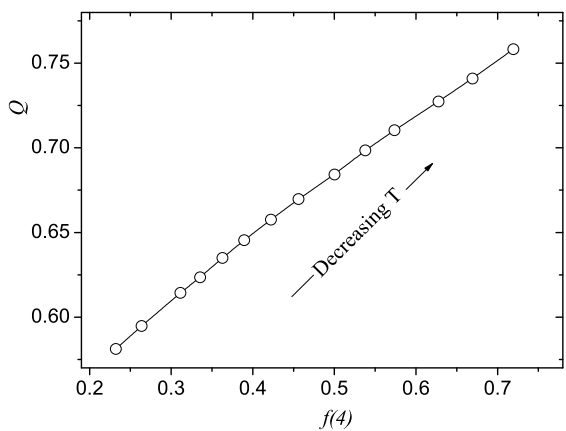}
\end{center}
\caption{ \textit{Q} vs $f(4)$. The change in $f(4)$ is higher than that of \textit{Q} in the range of temperatures studied. See the text for details.}
\label{figure03b}
\end{figure}

The $f(n)$ functions allow to obtain a more detailed picture of the structural orientational changes at shorter ranges between water molecules than the orientational order parameter. While a small change at 293 K occurs in the order map, the structures of two, three and four hydrogen bonds are alternated in importance when the temperature changes. The ability of $f(n)$ to more reliably describe the structure of the water occurs because the calculation of the hydrogen bond distributions includes the location of the hydrogen atoms, whereas \textit{Q} only quantifies the changes in the average angle between neighbor oxygen atoms. Although the behavior of $f(4)$ and \textit{Q} are correlated (see Fig. \ref{figure03b}), $f(4)$ shows a greater response to the temperature than \textit{Q}, indicating that the main change in the tetrahedral structure with the decrease of the temperature occurs  mainly in the orientation of the bonds between water molecules. The approximately linear correlation between both variables also indicates a similar dependence with the temperature ($\propto T^{-1}$).

\begin{figure}[h]
\begin{center}
\includegraphics*[width=7.4cm]{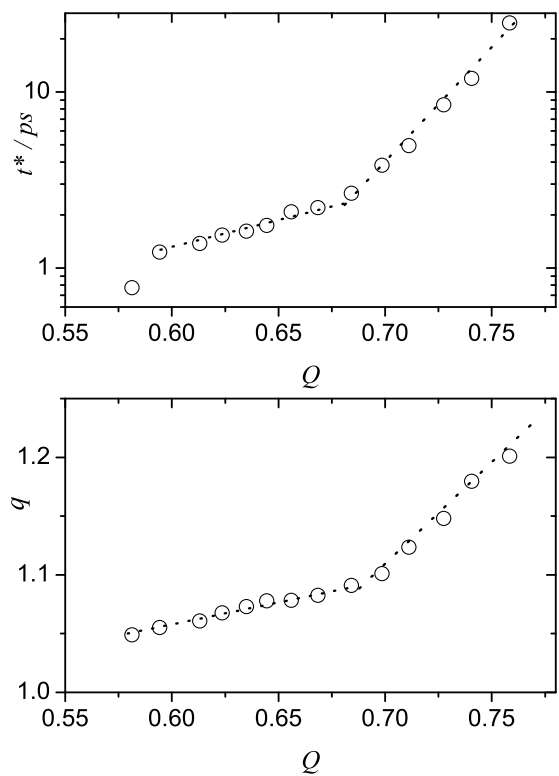}
\end{center}
\caption{(a) Semilog plot of $t^{*}$ vs. \textit{Q}. (b) $q$ vs. \textit{Q}. See the text for details.}
\label{figure05}
\end{figure}

Figure \ref{figure05} shows the behavior of the dynamical parameters $t^{*}$ and $q$ with \textit{Q}. The characteristic time $t^{*}$ has an exponential response to $Q \geq 0.58$ (T $\leq$ 360 K), but the slope of the semilog plot of $t^{*}$ vs. \textit{Q} increases significantly for $Q \geq 0.67$. A similar change occurs for $Q \geq 0.67$ in the linear correlation between $q$ and $Q$. Then, the values $Q \approx 0.67$ and $\tau \approx 1.1$ of the order map can be associated to changes in the dynamics of the system. The transition of $q\approx 1$ to $q>1$ indicates the increase of the probability of two water molecules remaining bonded by a hydrogen bond during an unusual long time, whereas the increase of $t^{*}$ is associated to the increase of the time during which the molecules remain in a subdiffusive regime.

However, only the analysis of the $f(n)$ functions reveals the structural modification that explains the structural and dynamic changes in the system. The changes in the increase of the order map, $t^{*}(Q)$ and $q(Q)$ occur below 293 K, in the range of temperatures in which prevail a structure of four hydrogen bonds in the system.

\section{Conclusions}

The molecular dynamic method allows to study the structure and dynamics of the SPC/E model of water in the range of temperatures from 213 K to 360 K. 

Lowering the temperature of the system from 360 K to 213 K, the number of water molecules in the first hydration layer decreases from $N \sim 5$ to $N \sim 4$, along with a decrease in size. The increase of the tetrahedral structure of the system is also characterized by a growth of the percentage of occurrence of four hydrogen bonds and the orientational order parameter $Q$. However, only the analysis of the behavior of the hydrogen bond distribution allows to deduce that, when a tetrahedral structure associated to the percentage of four hydrogen bonds predominates, the behavior of the dynamical variables $P(t)$ and $t^{*}$ show the occurrence of long lasting hydrogen bonds and \textit{caging} effect between the molecules of the system.

\begin{acknowledgements}
 I am grateful for the financial support by PICTO UNLPAM 2005 30807 and Facultad de Ciencias Exactas y Naturales (UNLPam).
\end{acknowledgements}

\end{document}